\documentclass[floatfix,aps,prl,twocolumn,showpacs,superscriptaddress,preprintnumbers]{revtex4-1}

\usepackage[english]{babel}
\usepackage{bm}    
\usepackage{epsfig}
\usepackage{amsbsy,amssymb}
\usepackage{graphicx}
\usepackage{array}
\usepackage{amsmath}
\usepackage{multirow}
\usepackage{natbib}
\usepackage{xspace}

\newcommand\new{\newcommand}         

\def\beq{\begin{equation}}
\def\eeq{\end{equation}}
\def\bea{\begin{eqnarray}}
\def\eea{\end{eqnarray}}
\newcommand{\bite}{\begin{itemize}}
\newcommand{\eite}{\end{itemize}}

\new{\proton}{\ensuremath{\mathrm{p}}\xspace}
\new{\Higgs}{\ensuremath{\mathrm{H}}\xspace}
\new{\Wboson}{\ensuremath{\mathrm{W}}\xspace}
\new{\Zboson}{\ensuremath{\mathrm{Z}}\xspace}
\new{\Wjjj}{\Wboson\!+\! 3\! jets\xspace}
\new{\Hj}{\Higgs\!+\! 1\! jet\xspace}
\new{\Hjp}{\Higgs\!+\! 1\! jet\phantom{s}\xspace}
\new{\Hjj}{\Higgs\!+\! 2\! jets\xspace}
\new{\Hjjj}{\Higgs\!+\! 3\! jets\xspace}
\new{\Hjjjj}{\Higgs\!+\! 4\! jets\xspace}
\new{\Hjets}{\Higgs\!+\! 1,\,2 and 3\! jets\xspace}
\new{\Hnj}{\Higgs\!+\! $n$\! jets\xspace}

\new{\eV}         {{\ifmmode {\mathrm{ eV}}\else ${\mathrm{ eV}}$\fi}}
\new{\MeV}        {{\ifmmode {\mathrm{ MeV}}\else ${\mathrm{ MeV}}$\fi}}
\new{\MeVc}       {{\ifmmode {\mathrm{ MeV}}/c\else ${\mathrm{ MeV}}/c$\fi}}
\new{\MeVcc}      {{\ifmmode {\mathrm{ MeV}}/c^2\else ${\mathrm{ MeV}}/c^2$\fi}}
\new{\GeV}        {{\ifmmode {\mathrm{ GeV}}\else ${\mathrm{ GeV}}$\fi}}
\new{\GeVc}       {{\ifmmode {\mathrm{ GeV}}/c\else ${\mathrm{GeV}}/c$\fi}}
\new{\GeVcc}      {{\ifmmode {\mathrm{ GeV}}/c^2\else ${\mathrm{GeV}}/c^2$\fi}}
\new{\TeV}        {{\ifmmode {\mathrm{ TeV}}\else ${\mathrm{ TeV}}$\fi}}

\new{\Mh}         {{\ifmmode M_{\mathrm{ H}}
                    \else $M_{\mathrm{H}}$\fi}}
\new{\Mz}         {{\ifmmode M_{\mathrm{Z}}
                    \else $M_{\mathrm{Z}}$\fi}}
\new{\Mzsq}       {{\ifmmode M^2_{\mathrm{ Z}}
                    \else $M^2_{\mathrm{Z}}$\fi}}
\new{\as}[1]      {{\ifmmode\alpha^{#1}_s
                    \else$\alpha^{#1}_s$\fi}}
\new{\asx}[1]      {{\ifmmode a^{#1}_s
                    \else $a^{#1}_s$\fi}}
\new{\asb}[1]     {{\ifmmode\overline{\alpha}^{#1}_s
                    \else $\overline{\alpha}^{#1}_s$\fi}}
\new{\asmz}       {{\ifmmode\alpha_s(\Mzsq)
                    \else $\alpha_s(\Mzsq)$\fi}}
\new{\lqcd}       {{\ifmmode\Lambda_{\mathrm{ QCD}}
                    \else $\Lambda_{\mathrm{ QCD}}$\fi}}

\def\Gosam{{\sc GoSam}\xspace}

\def\form{{\sc form}\xspace}

\def\Ninja{{\sc Ninja}\xspace}

\def\C++{{\sc c++}\xspace}

\def\POWHEG{{\sc POWHEG-BOX-V2}\xspace}
\def\MCFM{{\sc MCFM}\xspace}

\def\OneLoop{{\sc OneLoop}\xspace}
\def\Openloops{{\sc OpenLoops}\xspace}

\def\FastJet{{\sc FastJet}\xspace}
\def\Madgraph{{\sc Madgraph4}\xspace}

\new{\heft}{HEFT\xspace}
\new{\ftap}{FT$_{\mathrm{approx}}$\xspace}
\new{\mh}{m_{\mathrm{H}}}
\new{\mt}{m_{\mathrm{T}}}
\new{\pt}{p_{t}}
\new{\pth}{p_{t,\mathrm{H}}}
\new{\ptj}{p_{t,\mathrm{j}}}
\new{\ptji}{p_{t,\mathrm{j_i}}}
\new{\hthatprime}{\hat{H}^\prime_T}

\new{\HJ}{\texttt{HJ}\xspace}

\begin{document}

\title{NLO QCD corrections to Higgs boson plus jet production with full top-quark mass dependence}

\author{S.~P.~Jones}
\affiliation{Max-Planck-Institut f\"ur Physik, F\"ohringer Ring 6,
80805 M\"unchen, Germany}
\author{M.~Kerner}
\affiliation{Max-Planck-Institut f\"ur Physik, F\"ohringer Ring 6,
80805 M\"unchen, Germany}
\author{G.~Luisoni}
\affiliation{Max-Planck-Institut f\"ur Physik, F\"ohringer Ring 6,
80805 M\"unchen, Germany}


\preprint{MPP-2018-7}
\begin{abstract}
We present the next-to-leading order QCD corrections to the production
of a Higgs boson in association with one jet at the LHC including the
full top-quark mass dependence. The mass of the bottom quark is
neglected. The two-loop integrals appearing in the virtual
contribution are calculated numerically using the method of Sector
Decomposition.  We study the Higgs boson transverse momentum
distribution, focusing on the high $\pth$ region, where the top-quark
loop is resolved.  We find that the next-to-leading order QCD
corrections are large but that the ratio of the next-to-leading order
to leading order result is similar to that obtained by computing in
the limit of a large top-quark mass.
\end{abstract}



\maketitle

\section{Introduction}
\label{Sec:intro}

During Run I and early Run II of the LHC great progress has been made
in establishing many of the properties of the Higgs boson particle
discovered in 2012. Already the spin and CP properties are well
constrained and its couplings to the Standard Model (SM) weak vector
bosons and heavier fermions (top quarks and tau leptons) have been
measured \cite{Khachatryan:2016vau}. So far the measured properties
are consistent with the predictions of the SM.

Among the different channels for the production of a SM Higgs boson,
gluon fusion, which we consider here, is the mechanism yielding the
largest contribution. At lowest order in perturbation theory this
process is mediated by a closed loop of heavy quarks and
Next-to-Leading Order (NLO) QCD corrections therefore require the
computation of two-loop contributions. A Higgs Effective Field Theory
(\heft) was derived long ago~\cite{Wilczek:1977zn}, in which the heavy
quarks are integrated out and the Higgs boson couples directly to the
gluons. This allows the computations to be simplified considerably.

One interesting regime to consider is that of Higgs boson production
with a transverse momentum $\pth$ of the order of the top-quark mass,
$\mt$, or larger. Here the top quark loop is resolved and it becomes
possible to disentangle the SM contribution from effects of New
Physics. However, in this regime finite top-quark mass effects are not
negligible and the effective theory approximation becomes increasingly
poor. In other words, events in which the Higgs boson is recoiling
against one or more jets acquiring a large transverse momentum do not
fall into the validity range of the effective field theory
description. It is thus important to go beyond the \heft approximation
and include finite top-quark mass effects to obtain reliable
predictions in this kinematical range.

Within the \heft approximation corrections to inclusive Higgs boson
production are known to N$^3$LO QCD
accuracy~\cite{Anastasiou:2016cez}, whilst the fully differential
corrections for \Hj production are known to NNLO QCD
accuracy~\cite{Boughezal:2013uia,Chen:2014gva,Boughezal:2015dra,Boughezal:2015aha}.
Finite quark mass corrections to \Hj production have been known at LO
for a long time~\cite{Ellis:1987xu,Baur:1989cm} and LO results are
also known for the higher multiplicity processes
\Hjj~\cite{DelDuca:2001eu,DelDuca:2001fn} and recently
\Hjjj~\cite{Campanario:2013mga,Greiner:2016awe}.  The \heft results
for \Hj production have also been supplemented by an expansion in $1/\mt^2$ at
NLO QCD accuracy~\cite{Harlander:2012hf,Neumann:2014nha} and combined
with the exact Born and real corrections~\cite{Neumann:2016dny}.  They
were also included in multi-purpose Monte Carlo generators to produce
merged samples matched to parton
showers~\cite{Buschmann:2014sia,Hamilton:2015nsa,Frederix:2016cnl} and
used to improve the Higgs NNLO QCD transverse momentum distributions
in the \heft above the top-quark mass threshold~\cite{Chen:2016zka}.

One of the first major steps towards the computation of the full
two-loop NLO QCD virtual corrections was made in
Ref.~\cite{Bonciani:2016qxi}, where the planar master integrals were
computed analytically in the Euclidean region and shown to contain
elliptic integrals.  At the same time an expansion valid in the limit
of small bottom-quark mass allowed insight to be gained into the NLO QCD
effects due to nearly massless
quarks~\cite{Mueller:2015lrx,Melnikov:2016qoc,Melnikov:2017pgf,Lindert:2017pky}.
Very recently a NLO QCD result expanded in the regime where the Higgs
boson and top-quark masses are small, relevant for the description of
the Higgs boson transverse momentum distribution at large $\pth
\gtrsim 400$ \GeV, was also
studied~\cite{Kudashkin:2017skd,Lindert:2018iug}.

On the experimental side, recently the CMS collaboration has
considered events where the Higgs boson transverse momentum $\pth$ is
larger than 450~\GeV~\cite{Sirunyan:2017dgc}; a feat made possible
through the use of boosted techniques~\cite{Butterworth:2008iy}, which
allow the Higgs to be identified through its decay to bottom quarks.

In this letter we present the first NLO QCD computation of Higgs boson
production in association with one jet retaining the full top-quark
mass dependence.  In the following sections we present the
computational setup used for this calculation and selected
phenomenological results.

\section{Computational setup}
\label{sec:calc}

We compute using conventional dimensional regularization (CDR) with
$d=4-2 \epsilon$. The top-quark mass is renormalized in the on-shell
scheme and the QCD coupling and gluon wave-function in the
$\overline{\mathrm{MS}}$ scheme with $n_f = 5$ light quarks, with the
top quark loops subtracted at zero momentum.  The top-quark mass
renormalization is performed by inserting the mass counterterm into
the heavy quark propagators. Alternatively, the mass renormalization
can be calculated by taking the derivative of the one-loop amplitude
with respect to $\mt$. We have used both methods as a cross-check.

The sampling of the phase-space generator has been adapted to generate
a nearly uniform distribution of points in $\pth$.

\subsection{Born and Real radiation}

The computation of everything but the virtual amplitudes is performed
within the \POWHEG framework~\cite{Alioli:2010xd}, taking advantage of
the existing \HJ generator~\cite{Campbell:2012am} for \heft, in which
the Born and real radiation amplitudes are computed using
\Madgraph~\cite{Stelzer:1994ta,Alwall:2007st} and the virtual
amplitudes are taken from \MCFM~\cite{Campbell:2010cz}. The
subtraction of the infra-red divergences is performed using
FKS~\cite{Frixione:1995ms}.

We supplemented this code with the analytical Born amplitudes with
full top-quark mass dependence from Ref.~\cite{Baur:1989cm}, whereas
the one-loop real radiation contribution was generated with
\Gosam~\cite{Cullen:2011ac,Cullen:2014yla} using the
BLHA~\cite{Binoth:2010xt,Alioli:2013nda} interface developed in
Ref.~\cite{Luisoni:2013kna}. For the purpose of this computation
\Gosam has been improved such that it now automatically switches to
quadruple precision in regions where the amplitude becomes unstable
due to one of the final state partons becoming soft or collinear to
another parton. The amplitudes generated by \Gosam are computed at run
time with
\Ninja~\cite{Mastrolia:2012bu,vanDeurzen:2013saa,Peraro:2014cba} using
the \texttt{quadninja} feature.  The scalar one-loop integrals are
computed with the \OneLoop~\cite{vanHameren:2010cp} integral library.
As a consistency check the virtual amplitudes in \heft were cross
checked with \Gosam, whereas the real radiation amplitudes in the full
theory were compared against \Openloops~\cite{Cascioli:2011va}.

\subsection{The virtual amplitude}

The Lorentz structure of the $H \rightarrow q\bar{q}g$ and $H
\rightarrow ggg$ partonic amplitudes can be decomposed, after imposing
parity conservation, transversality of the gluon polarization vectors
and the Ward identity, in terms of 2 and 4 tensor structures
respectively, see for example Ref.~\cite{Gehrmann:2011aa}. This
decomposition is not unique. For the $H \rightarrow ggg$ amplitude we
follow Ref.~\cite{Boggia:2017hyq} and choose to decompose it such that
three of the form factors (which multiply the tensor structures) are
related by cyclic permutations of the external gluon momenta whilst
the fourth is itself invariant under such permutations. For the $H
\rightarrow q\bar{q}g$ amplitude our decomposition is chosen such that
the form factors are related by interchanging the external quark and
anti-quark momenta.  We separately compute all form factors and use
these symmetries as a cross-check of our result.

In order to compute the amplitudes we closely follow the method of
Refs.~\cite{Borowka:2016ehy,Borowka:2016ypz}.  We construct projection
operators for each of the form factors and contract them with the
amplitude omitting external spinors and polarization vectors.  This
procedure allows us to write the amplitude in terms of scalar
integrals.

The Feynman diagrams contributing to the two-loop virtual amplitude
are generated using \textsc{Qgraf}~\cite{Nogueira:1991ex} and further
processed using \textsc{Reduze}~\cite{vonManteuffel:2012np},
\textsc{Ginac}~\cite{Bauer:2000cp}, \textsc{Fermat}~\cite{fermat}, and
\textsc{Mathematica}.  We cross checked the amplitudes with
expressions obtained from a two-loop extension of \Gosam, which uses
\textsc{Qgraf} and \form~\cite{Vermaseren:2000nd,Kuipers:2012rf}.  The
integrals appearing in the amplitude are reduced to master integrals
using a customized version of the program \textsc{Reduze}.  To
simplify the numerical evaluation we choose a quasi-finite basis of
master integrals~\cite{vonManteuffel:2014qoa}.  The resulting
integrals are calculated numerically using
\textsc{SecDec}~\cite{Borowka:2015mxa,Borowka:2017idc}. For the 
numerical integration we use a quasi-Monte Carlo method based on a rank-one
lattice rule~\cite{Li:2015foa,QMCActaNumerica,nuyens2006fast}.
Neglecting crossings we evaluate a total of 102 planar and 18 non-planar 
two-loop integrals.

The Higgs boson mass is set to $\mh=125~\GeV$ and the top-quark mass
is chosen such that $\mh^{2}/\mt^{2}=12/23$, which means that
$\mt=173.055~\GeV$. Fixing the ratio of the Higgs boson to top-quark
mass allows us to reduce by one the number of independent scales
appearing in the two-loop virtual amplitudes. This simplifies the
integral reduction and the form of the resulting reduced amplitude.

We subtract the infra-red and collinear poles of the virtual amplitude
to obtain the finite part of the virtual amplitude
$\mathcal{V}_\mathrm{fin}$ as required in the \POWHEG
framework~\cite{Heinrich:2017kxx}.  The IR subtraction procedure
requires the one-loop amplitudes up to order $\epsilon^2$, which we
compute numerically using the same procedure as for the two-loop
amplitudes.

\section{Phenomenology}
\label{sec:result}

In this section we present results for \Hj production at the LHC at a
center-of-mass energy of 13~\TeV. Jets were clustered using the {\tt
  anti-kt} jet algorithm implemented in
\FastJet~\cite{Cacciari:2005hq,Cacciari:2008gp,Cacciari:2011ma} with a
radial distance of $R=0.4$ and requiring a minimum transverse momentum
of $\ptj>30$~\GeV. We used the
PDF4LHC15{\tt\_}nlo{\tt\_}30{\tt\_}pdfas~\cite{Butterworth:2015oua,CT14,MMHT14,NNPDF}
set interfaced through LHAPDF~\cite{Buckley:2014ana} for both LO and
NLO predictions, and fixed the default value of factorization and
renormalization scales $\mu_{F}$ and $\mu_{R}$ to $H_T/2$, defined as
\begin{align}\label{eq:scale}
  \frac{H_T}{2}=\frac{1}{2}\Big(\sqrt{\mh^2+\pth^2}+\sum_{i}\left|p_{t,i}\right|\Big)\,,
\end{align}
where the sum runs over all final state partons $i$. This scale is
known to give a good convergence of the perturbative expansion and
stable differential $K$-factors (ratio of NLO to LO predictions) in
the effective theory~\cite{Greiner:2015jha}. To estimate the
theoretical uncertainty we vary independently $\mu_{F}$ and $\mu_{R}$
by factors of $0.5$ and $2$, and exclude the opposite variations. The
total uncertainty is taken to be the envelope of this 7-point
variation.

To better highlight the differences arising from the two-loop massive
contributions, we compare the new results with full top-quark mass
dependence, which we label as ``full theory result'' or simply
``full'' in the following, to two different approximations. In
addition to predictions in the effective theory, which are referred to
as \heft in the following, we show results in which everything but the
virtual amplitudes is computed with full top-quark mass dependence. In
this latter case only the virtual contribution is computed in the
effective field theory and reweighted by the full theory Born
amplitude for each phase space point. Following
Ref.~\cite{Maltoni:2014eza} we call this prediction ``approximated
full theory'' and label it as \ftap from now on.

We start by presenting the total cross sections, which are reported in
Table~\ref{tab:totxs}. For comparison we present results also for the
\heft and \ftap approximations.

\begin{table}[h]
\begin{tabular}{ l  c  c }
\hline
\hline
{\sc Theory} & LO [pb] & NLO [pb]\\
\hline
\heft: & $\sigma_{\mathrm{LO}} = 8.22^{+3.17}_{-2.15}$ & $\sigma_{\mathrm{NLO}} = 13.53^{+2.19}_{-2.04}$ \phantom{\Big|}\\
\ftap: & $\sigma_{\mathrm{LO}} = 8.57^{+3.31}_{-2.24}$ & $\sigma_{\mathrm{NLO}} = 14.06^{+2.17}_{-2.25}$ \phantom{\Big|}\\
Full:  & $\sigma_{\mathrm{LO}} = 8.57^{+3.31}_{-2.24}$ & $\sigma_{\mathrm{NLO}} = 14.19(7)^{+2.29}_{-2.23}$ \phantom{\Big|}\\
\hline
\hline
\end{tabular}
\caption{Total cross sections at LO and NLO in the \heft and \ftap approximations and with full top-quark mass dependence. The upper and lower values due to scale variation are also shown. More details can be found in the text.}
\label{tab:totxs}
\end{table}

Together with the prediction obtained with the central scale defined
according to Eq.~(\ref{eq:scale}) we show the upper and lower values
obtained by varying the scales. While at LO the top-quark mass effects
lead to an increase of $4.3\%$, at NLO this is enhanced to $4.9\%$
when compared to the \heft approximation.
Comparing the full theory result with the \ftap
result we obtain an increase of $1\%$.  It is important to keep in mind that
when taking into account massive bottom-quark loop contributions, the
interference effects are sizable and cancel to a large extent the
increase in the total cross section observed here between the \heft
and the full theory results (see e.g. the results in
Ref.~\cite{Greiner:2016awe}). At LO the bottom-quark
mass effects are of the order of $2\%$ or smaller above the top
quark threshold.

\begin{figure}[t!]
  \centering
  \includegraphics[width=0.49\textwidth]{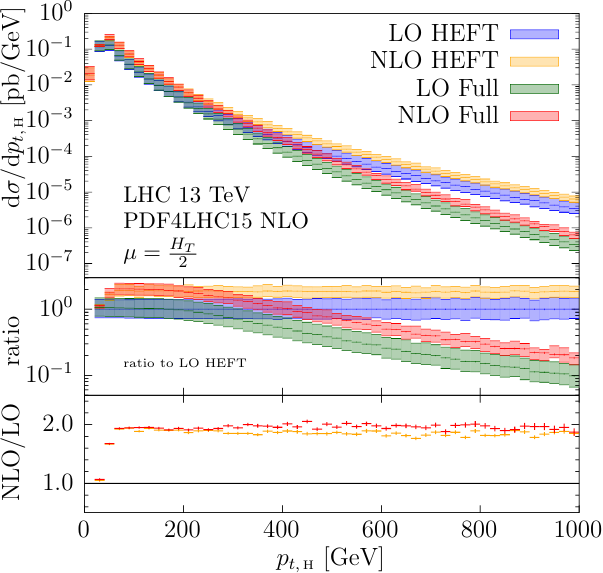}
  \caption{\label{fig:pth_fullheft}%
    Higgs boson transverse momentum spectrum at LO and NLO in QCD in
    \heft and with full top-quark mass dependence. The upper panel
    shows the differential cross sections, in the middle panel we
    normalize all distributions to the LO \heft prediction and in the
    lower panel we show the differential $K$-factors for both the
    \heft and the full theory distributions. More details can be found
    in the text.}
\end{figure}

Considering more differential observables, it is well known that very
significant effects due to resolving the top-quark loop are displayed
by the Higgs boson transverse momentum distribution, which is softened
for larger values of $\pth$ by the full top-quark mass dependence.  By
considering the high energy limit of a point-like gluon-gluon Higgs
interaction and one mediated via a quark loop it is possible to derive
the scaling of the squared transverse momentum distribution
$d\sigma/d\pth^2$~\cite{Forte:2015gve,Caola:2016upw}, which drops as
$(\pth^2)^{-1}$ in the effective theory, and goes instead as
$(\pth^2)^{-2}$ in the full theory. This fact was shown to hold
numerically at LO for up to three jets in
Ref.~\cite{Greiner:2016awe}. It is interesting to verify this also
after NLO QCD corrections are applied. To do so, in
Figure~\ref{fig:pth_fullheft} we show the transverse momentum spectrum
of the Higgs boson at LO and NLO in the \heft approximation and with
the full top-quark mass dependence.

In the upper panel we display each differential distribution with the
theory uncertainty band originating from scale variation. To highlight
the different scaling in $\pth$, in the middle panel we normalize all
the distributions to the LO curve in the effective theory. It is thus
possible to see that for low transverse momenta the full theory
predictions overshoot slightly the effective theory ones. For
$\pth>200~\GeV$ the two predictions start deviating more
substantially. At LO the two uncertainty bands do not overlap any more
above $400~\GeV$, whereas at NLO this happens already around
$340~\GeV$ due to reduction of the uncertainty at this order. The
logarithmic scale also allows to see that the relative scaling
behavior within the two theory descriptions is preserved between LO
and NLO. The curves in the lowest panel of
Figure~\ref{fig:pth_fullheft} show the differential $K$-factor in
\heft and in the full theory. In both cases above $150~\GeV$ they
become very stable and amount to about $1.85$ and $1.95$
respectively. Thus the NLO corrections are large also in the full
theory. This broadly agrees with the conclusions of
Ref.~\cite{Lindert:2018iug}, where the expanded result showed a
similar enhancement of the $K$-factor by about $6\%$ in the tail
compared to the \heft.

\begin{figure}[t!]
  \centering
  \includegraphics[width=0.49\textwidth]{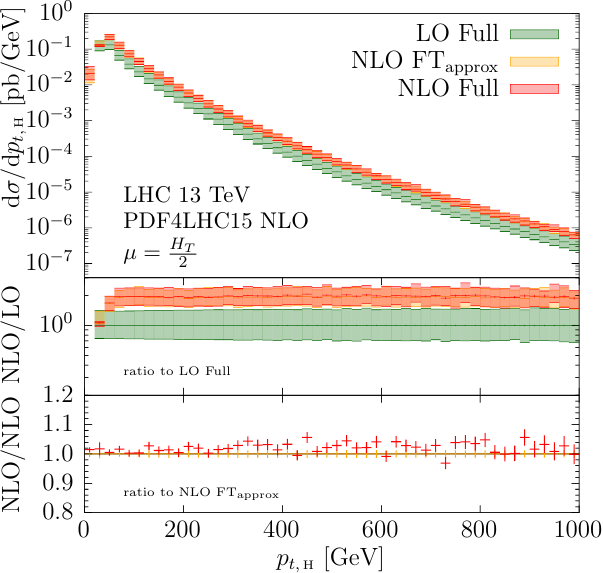}
  \caption{\label{fig:pth_fullftap}%
    Higgs boson transverse momentum spectrum at LO and NLO in QCD with
    full top-quark mass dependence compared with the NLO predictions
    in \ftap. The upper panel shows the differential cross sections,
    in the middle panel we normalize all distributions to the LO
    prediction and in the lower panel we show the differential ratio
    between the NLO \ftap predictions and the full theory ones. More
    details can be found in the text.}
\end{figure}

To conclude this section we compare the new predictions for the Higgs
boson transverse momentum with the one in \ftap. At LO the two
predictions are identical by construction, it is however interesting
to check how good \ftap can reproduce the full theory results. In the
main panel of Figure~\ref{fig:pth_fullftap} we plot the three
curves. To highlight better the differences among the two predictions,
in the middle panel we normalize the distributions to the LO
prediction. This allows us to compare the two differential $K$-factors,
which behave very similarly over the full kinematical range. 
In order to quantify the difference between the two
predictions, in the lower panel we display the ratio of the Full NLO
curve to the \ftap NLO curve showing an increase at the 1-2\% level of
the Full NLO compared to the \ftap prediction.

\section{Conclusions and Outlook}
\label{sec:conc}

In this letter we have presented for the first time NLO QCD
corrections to Higgs boson plus jet production retaining the full
top-quark mass dependence. We observe that the size of the NLO
corrections is large but, for our choice of the renormalization and
factorization scale, the $K$-factor is approximately constant above
the top-quark threshold. Compared to \ftap predictions, the full
two-loop contribution enhances the NLO predictions by about $1\%$ at
the level of the total cross section, with nearly no dependence 
on the transverse momentum of the Higgs boson.
Despite a completely different $\pt$ scaling, the
$K$-factors in the \heft and in the full theory behave in a very
similar way.

The result removes the theoretical uncertainty on differential \Hj
distributions due to the unknown top mass corrections at NLO in
QCD. Besides the transverse momentum distribution, shown here, this
calculation enables accurate predictions to be made also for other
observables where the top-quark mass effects may play a significant
role.

As the experimental precision at the LHC improves in the coming years,
this result aids the study of the Higgs boson properties also in
boosted regimes. Providing a more accurate theoretical description of
the Higgs boson production at large transverse momentum will be
helpful not only for unraveling the details of the electroweak
symmetry breaking mechanism, but also in the search for indirect signs of new
physics.

\bigskip

\begin{acknowledgments}
  \noindent{\it Acknowledgments}\\
We would like to thank Gudrun Heinrich for encouraging us to undertake
this computation and also for numerous discussions and comments on the
manuscript. We thank Nigel Glover and Hjalte Frellesvig for
interesting discussions and for their insightful comments on the
Lorentz structure of the amplitude. We thank Tiziano Peraro for
helping us in the upgrade of the \Ninja interface and Sophia Borowka
for interesting discussions. We thank Xuan Chen for detailed cross checks 
  of the real-radiation contributions.
  This research was
supported by the Munich Institute for Astro- and Particle Physics
(MIAPP) of the DFG cluster of excellence ``Origin and Structure of the
Universe''.  SPJ was supported by the Research Executive Agency (REA)
of the European Union under the Grant Agreement {PITN-GA}2012316704
(HiggsTools) during part of this work.  We gratefully acknowledge
support and resources provided by the Max Planck Computing and Data
Facility (MPCDF).
\end{acknowledgments}


\bibliographystyle{apsrev}
\bibliography{references.bib}

\end{document}